\documentstyle[12pt]{article}

\title{Dynamics of strong-field photoionization
in two dimensions for short-range binding potentials}

\author{J. Matulewski\footnote{e-mail: jacek@phys.uni.torun.pl},
A. Raczy\'nski and J. Zaremba\\
Instytut Fizyki,Uniwersytet Miko\l aja Kopernika,\\
ul. Grudzi\c{a}dzka 5, 87-100 Toru\'n, Poland}

\begin{document}
\maketitle

\begin{abstract}
Photodetachment in an ultrastrong laser field in two spatial
dimensions is investigated {\em ab initio}. A qualitative behavior
of the packet for a short-range binding potential is contrasted
with that for a soft-core potential, in particular dynamical
effects due to a rescattering of  fragments separated from the
main packet are demonstrated.
\end{abstract}

\section{Introduction}
The problem of ultrastrong-field photoionization has been the
subject of studies in quantum optics for the last fifteen years
(for a recent review see Ref. \cite{1}). Because of enormous
technical difficulties most of \emph{ab initio} calculations were
performed for one dimensional models. It is well known that if the
electric field amplitude $\varepsilon_{0}$ grows in the region of
a few atomic units (for a typical frequency $\omega = 1$ a.u. and
a binding potential supporting an initial state of energy of order
of 1 a.u.), the ionization rate need not grow monotonically with
the field and the population can be regarded as trapped in a
Kramers-Henneberger (KH) well \cite{2}. At a field amplitude of
about 10 a.u. almost no probability leaks from the neighborhood of
the nucleus. Unfortunately, one-dimensional models fail to
represent important features of real systems; in particular they
can imitate only an interaction with a linearly polarized field in
the dipole approximation. Thus investigating magnetic interactions
or light polarization effects requires a generalization to models
with more spatial dimensions \cite{3}. Investigations performed in
two dimensions for long-range binding potentials predict in
particular that the stabilization at field intensities higher than
10 a.u. is destroyed by a nondipole kinetic effect – the so-called
magnetic drift of the packet as a whole away from the region of
the nucleus.

The aim of this paper is to study photodetachment in superintense
fields (such that the Schr\"odinger equation is still valid) in
two dimensions in the situation which has not yet been
investigated in a few so far existing papers \cite{3,4,8}; in
comparison with other papers the long-range atomic potential is
replaced by a short-range potential well (rectangular). The
details of the dynamics in our case include some special effects,
an interpretation of which is necessary for a complete
understanding of the process.

A new element in comparison with earlier papers \cite{3} is that
the well potential (especially if it has deep eigenstates), due to
its steep edges, causes tearing of the oscillating wave packet
into fragments, even in such strong laser field intensities as 15
- 20 a.u. Because of the wavepacket tearing, the classical model
does not allow to satisfactorily interpret all numerical results.
In particular satellite wavepacket's fragments detached at
different times and being of a significant size can be rescattered
in their oscillatory motion and interfere, with the details of the
process depending on the shape of the laser pulse, in particular
on the position of the classical packet's turning points. The
phase dependence of the dynamics is also discussed.
\\

\section{The binding potential and the shape of the pulse}
There are two types of model atomic potentials commonly used in
the studies on the interaction of atomic systems with superstrong
fields in one-dimension: a soft core Coulomb potential proposed by
Su, Javanainen and Eberly in a series of papers \cite{4} and a
short-range potential of a negative ion, modeled, e.g., by a
rectangular well. The main difference is that the former potential
is smooth and supports an infinite number of discrete states while
the latter one has two points of discontinuity and a few bound
states. In the latter case the packet has less chances to remain a
connected structure (i.e. it is more exposed to tearing) and
cannot be dynamically stabilized, i.e. trapped in the series of
the Rydberg states.

While two-dimensional studies on the strong-field stabilization of
models with a soft core potential have already been reported (see
\cite{1}, \cite{5}), in this paper we concentrate on special
effects due to a potential well, which we assume to be radial,
i.e. we take
\begin{equation}
V(\vec{r})=-V_{0}\theta(a-r),
\end{equation}
where $r=\sqrt{x^{2}+y^{2}}\label{eq1}$. The width $a$ and the
depth $V_{0}$ of the well are chosen so that there is only a
single (initial) bound state.
\\

The choice of the pulse shape is also very important for the
dynamics of the ionization process. The stabilization is possible
only if the pulse does not lead to a fast drift of the electron,
i.e. the classical trajectory of the electron in the absence of
the binding potential remains close to the nucleus. The results
presented below are obtained for rectangular cosinusoidal pulses
which satisfy this condition. For a pulse propagating along the
$y$ axis and with $x$ polarization we thus take the field
amplitude
$\vec{\varepsilon}(y,t)=\hat{x}\varepsilon_{0}\cos(ky-\omega t)$
or the vector potential
$\vec{A}(y,t)=-\hat{x}\frac{\varepsilon_{0}}{\omega}\sin(ky-\omega
t)$. The continuous pulses often adopted in other papers (e.g.
those with trapezoid envelope) may imply complicated switch-on
effects, especially if no dipole approximation is made. For
example they may give raise to a field component which is not an
electromagnetic wave. Moreover, calculating the vector potential
by a time integral of the electric field introduces ambiguities
reflecting various possible experimental realizations. We have
checked that the effect we describe below appears also in a
modified version for a trapezoid envelope.

The cosine pulse with any initial phase has an important feature
in contradistinction to other pulses usually assumed in this
context, namely the turning points are shifted asymmetrically with
respect to the potential well, with one of them located at the
well, i.e. at the initial position of the wave packet. If
$\varepsilon(t)=\varepsilon_{0}\cos(\omega t)$, the electron free
of the binding potential would, in one dimension, oscillate in the
range $(-2\varepsilon_{0}/\omega^{2} , 0)$, which means that the
right turning point occurs at the well.
\\

Strong field ionization is very often described in terms of the
Kramers-Henneberger (KH) well. In Kramer's frame an unbounded
electron stays at rest and experiences the interaction of the
oscillating nucleus. The KH potential well is just this
interaction averaged over the oscillation period or, in other
words, the zero-th term of the Fourier series of the interaction.
In the initial stage of ionization the electron's wavepacket
spreads out until it has reached the size of the KH well. Further
changes of the packet's shape as well as its slow drift may be
explained as due to higher terms of the Fourier expansion.

For the cosinusoidal pulse the electron is initially located close
to one of the two minima of the KH well, in contradistinction to
the case of other pulses studied in this context, for which its
initial position occurs in the middle.

In the present work we will demonstrate how the results of
one-dimensional computations must be completed and developed in
the case of two spatial dimensions. In particular the Kramers
frame and the corresponding KH well are now introduced as if there
were no movement in the direction perpendicular to the electric
field
$\vec{\varepsilon}(\vec{r},t)=\hat{x}\varepsilon_{0}\cos(ky-\omega
t)$: the transformation accounts for the electron movement
$X(t)=\frac{\varepsilon_{0}}{\omega^2}(1-\cos(\omega t))$, while
its slow drift in the $x$ direction and/or its throwing away of
small parts in both directions are due to the higher terms of the
Fourier expansion. The evolution in the $y$ direction includes
also magnetic effects seen as throwing away of small pieces rather
than a drift of the packet as a whole. This new aspect of the
two-dimensional ionization process has not been observed in other
works on strong-field ionization because they were dealing with
soft-core binding potentials, in contradistinction to short-range
potentials considered in this paper.
\\

The KH well has the form

\begin{equation}
V_{KH}(x,y;\varepsilon_{0})=\frac{1}{T}\int_{t}^{t+T}V(x-X(\tau),y)d\tau
, \label{eq2}
\end{equation}

where $T=2\pi/\omega$. The analytical form of the above KH well is
\begin{equation}
V_{KH}(x,y;\varepsilon_{0})=-\frac{V_{0}}{\pi}
\Theta(r^2-y^2)(\arcsin(\alpha_+)-\arcsin(\alpha_-)), \label{eq3}
\end{equation}

where

\begin{equation}
\alpha_{\pm}=
\left\{
\begin{array}{ll}
1, & x<\pm \sqrt{a^2-y^2} \\
1\pm\frac{\omega^2}{\varepsilon_0}(\sqrt{a^2-y^2}\mp x) & \\
-1, & x>2\pm \sqrt{a^2-y^2} \\
\end{array}.
\right.
\label{eq4}
\end{equation}

The KH potential for $\varepsilon_0=15$ a.u., $\omega=1$ a.u. and
for the radial well (Eq. (\ref{eq1}) with $a=1$ a.u., $V_0=2$
a.u., a single bound state) and its two eigenstates are shown in
Fig. \ref{fig1}.
\\

\section{Quantum numerical results of {\em ab initio}
simulations and classical predictions}
Because we want to include nondipole effects we write the
atom-field interaction in the velocity gauge. The vector potential
$\hat{x}A(ky-\omega t)$ is polarized along \emph{x} axis and the
Hamiltonian reads

\begin{equation}
\hat{H}(\vec{r},t)=\frac{(\hat{p}+\hat{x}A(ky-\omega
t))^2}{2}+V(\vec{r}), \label{eq5}
\end{equation}
where the radial potential is given by Eq. (\ref{eq1}).
\\

We have solved this equation both with and without the dipole
approximation using \emph{the alternating direction implicit}
method (ADI) \cite{6}, which allows one to use tridiagonal sets of
equations which can be solved easily and fast. The main difficulty
in the two-dimensional calculation is not the number of operations
to be performed by a processor but rather a large memory necessary
to store the results. The size of the accessible memory sets a
limit to the size of the spatial grid used in the calculation. We
performed our calculation on the $2048\times2048$ grid with the
step $\Delta x$ close to 0.1 a.u. The space range is from $-100$
a.u. to $+100$ a.u. in both directions.
\\

As mentioned above the particular form of the vector potential was

\begin{equation}
\vec{A}(y,t)=-\hat{x}\frac{\varepsilon_0}{\omega}\sin(ky-\omega
t), \label{eq6}
\end{equation}
with $\omega = 1$ a.u. and $\varepsilon_0$ equal to 15 and 20 a.u.
Larger values of the vector potential amplitude require a
relativistic approach.

At the beginning of each cycle the left minimum of the KH well
occurs close to the packet's initial position in the well; in fact
Eq. (\ref{eq3}) yields a shift of the maximum of the wavepacket
trapped by the KH well with respect to the initial wavepacket's
maximum by about 1 a.u. The latter is localized in the KH well's
left half and is almost equally divided between the two discrete
states. The energy difference between the KH eigenstates is small,
so we do not observe any beats, which would occur in a much longer
time scale than that considered here. The shift of the minima of
the two wells is reflected on the dynamics of the initial state
population (see Fig. \ref{fig2}). The splitting of each peak can
be explained by the fact that the maximum of the trapped packet
passes by the original well twice at the beginning of each period,
with small velocities of opposite signs. This can be treated as an
evidence of applicability of KH approach in this range of
parameters.
\\

The rest of the initial wavepacket, i.e. the part which has not
been trapped in the KH well as well as the part initially trapped
but later released due to higher terms of the Fourier expansion,
is thrown away in the form of concentric rings (see Fig.
\ref{fig3}) which oscillate as a whole along the \emph{x} axis in
the rhythm of the field. In the dipole approximation neither turn
of the $y$ axis is distinguished but if no dipole approximation is
made most of probability will drift towards the positive direction
of the \emph{y} axis due to the magnetic force, according to the
classical considerations presented below. The additional
structures visible in Fig. \ref{fig3} will be discussed later.
\\

The vector potential given by Eq. (\ref{eq6}) corresponds to the
electric and magnetic fields:

\begin{eqnarray}
\vec{E}(y,t)=\hat{x}E(y,t)=\hat{x}\varepsilon_0\cos (ky-\omega
t),\\ \vec{B}(y,t)=-\hat{z}B(y,t)=-\hat{z}B_0\cos (ky-\omega t),
&{\rm where} & \vec{k}=\hat{y}k=\hat{y}\frac{y}{c}. \label{eq7}
\end{eqnarray}
Then the Lorentz force is

\begin{equation}
\vec{F}=-(\vec{E}+\vec{v}\times\vec{B})=-[E-v_yB,v_xB,0]
\label{eq8}
\end{equation}
One can write the classical equations of motion for an electron in
the electromagnetic field

\begin{equation}
\left\{
\begin{array}{ll}
\ddot{x}(t)=-\varepsilon_0(1-\frac{\dot{y}(t)}{c})\cos(ky(t)-\omega t),\\
\ddot{y}(t)=-\frac{\varepsilon_0}{c}\dot{x}(t)\cos(ky(t)-\omega t),\\
\end{array}
\right. \label{eq9}
\end{equation}
\\
which can be easily solved numerically. The solution of these
equations is presented in Fig. \ref{fig4}. The upper plot shows
fast oscillations with the frequency $\omega$ and the same
amplitude as for the corresponding one-dimensional solution in the
dipole approximation. The lower plot shows the magnetic drift and
the oscillations of the frequency $2\omega$. The velocity of this
drift is proportional to the square of electric field amplitude
$\varepsilon_0$. The results can be easily understood if we
simplify equations (\ref{eq9}) neglecting the magnetic component
of the Lorentz force in the \emph{x} direction and neglecting at
this stage the \emph{y} dependence of the field. So we can write
the simplified equations of motion

\begin{equation}
\left\{
\begin{array}{ll}
\ddot{x}(t)=-\varepsilon_0\cos(\omega t),
\dot{x}(t)=-\frac{\varepsilon_0}{\omega}\sin(\omega t),\\
\ddot{y}(t)=-\frac{\varepsilon_0}{c}\dot{x}(t)\cos(\omega t)=
\frac{\varepsilon_0^2}{2\omega c}\sin(2\omega t),\\
\end{array}
\right. \label{eq10}
\end{equation}
\\
with the electron initially resting at $x=y=0$.

This approximation corresponds to the series expansion of vector
potential $A_0\sin(ky-\omega t)$ around $\omega t$ and leaving
only the lowest term. The magnetic drift is directed towards the
positive values of \emph{y}, with oscillations of frequency
$2\omega$ imposed on it. The average velocity in the \emph{y}
direction is proportional to the square of the field amplitude.
\\

\section{Wavepacket tearing and its consequences}
The \emph{ab initio} solutions of the Schr\"{o}dinger equation
with the Hamiltonian (\ref{eq5}) and the laser field defined by
(\ref{eq6}) demonstrate that the mean values of the operators $x$
and $y$ (see Fig. \ref{fig5}) look very similar to the classical
predictions (Fig. \ref{fig4}), as could be expected from the
Ehrenfest theorem.
\\

The wavefunctions (see Figs \ref{fig3}, \ref{fig6a} and
\ref{fig6b}) look quite different from those shown in Refs
\cite{5} and \cite{4}, because here a deep potential well is used
instead of a soft-core atom. In spite of the strength of the laser
field, the wave function does not behave typically for the
over-the barrier (OTB) ionization regime, in which the packet
behaves essentially as a connected structure. Parts of the packet
are thrown away in the form of two sets of elliptic rings with
centers corresponding to the positions of both minima of the KH
well. Those oval structures carry a significant portion of the
probability of finding the electron. In the dipole approximation a
new ring appears once a period (Fig. \ref{fig6b}). If no dipole
approximation is made, the rings appear twice a period with each
second ring being due only to the magnetic force oscillating with
the doubled frequency (cf. Eq. (\ref{eq10})) - see the inset of
Fig. \ref{fig6a}. The torn-off structures are thrown preferably
towards positive $y$ due to the magnetic interaction. A comparison
of the heights of the main peak in Figs \ref{fig6a} and
\ref{fig6b} shows that making the dipole approximation leads to a
significant underestimation of the ionization rate.

In Fig. \ref{fig5}, similarly as in Fig. \ref{fig4}, we can see
fast oscillations in the \emph{x} direction in the upper plot and
the magnetic drift in the \emph{y} direction in lower one. A slow
drift, similar to that obtained in \cite{7}, is also visible at
the graph of $\langle x\rangle$.
\\

\section{Rescattering of the liberated parts of the wavepacket}

For the cosine rectangular pulse the wavepacket oscillates from 0
to $-2\varepsilon_0/\omega^2$ along the \emph{x}-axis like in the
OTB regime. It is well known that an irreversible ionization of
the packet in the form consisting in throwing away pieces of the
packet can occur only if both the electric field and the binding
potential are present and the packet has a small velocity when
passing by the well. Those conditions are satisfied during a
significant part of each cycle if the phase of the field is chosen
so that the pulse is cosinusoidal. Then one of the wavepacket's
turning points is located at the well so once a cycle an electron
has the possibility to be liberated. In the case of a sine pulse
with a trapezoid envelope (not shown here) there is no opportunity
for any packet tearing in such strong laser fields until the
wavepacket is wide enough to cover the range of classical
oscillations (i.e. the width of KH well). In this case rings
appear a few cycles later then in the case of cosine pulse.

As described above, for a cosine pulse at the end of each cycle,
when the packet is stopping above the well, a part of the packet
is thrown away from the well and forms a ring around the main part
of the packet. Thus the wavefunction takes the shape of rings
around the main maximum, oscillating as a whole in the $x$
direction. After a few periods the first oval structure reaches a
radius of order of $2\varepsilon_0/\omega^2$, so there is some
probability of finding the electron at the well at the moment at
which the classical turning point occurs at
$-2\varepsilon_0/\omega^2$, i.e. after an odd multiple of half a
cycle. Tearing this secondary cloud causes a creation of a second
family of rings, which are centered at the distance
$2\varepsilon_0/\omega^2$ from the center of the first family of
rings and obviously carry a smaller portion of the probability. \\

The most interesting effect is an interference of both families of
the rings giving rise to an additional structure between two
turning points and to a complicated shape of the wavefunction at
larger $x$ (see Fig. \ref{fig3} (the wavefunction after $10T$
computed in dipole approximation) and Figs \ref{fig6a} and
\ref{fig6b}). This interference effect has no magnetic character
and occurs similarly both in the general case and in the dipole
approximation.

This new type of rescattering can be even easier understood using
the KH model. In the Kramers frame, i.e. in the frame oscillating
like an unbound electron, the two wavepacket's turning points
correspond to the two turning points of moving potential and,
after time averaging, approximately to the minima of the KH well
(for cosine pulse: $x=0$ and $x=+2\varepsilon_0/\omega^2$).
Because for the field magnitude used in this paper the KH well has
two states with close energies (see Fig. \ref{fig1}) the electron
is located in such a superposition of those states that only the
left minimum is significantly populated. The higher terms of the
Fourier expansion, characterized by large frequencies but rather
small amplitudes, cause an ionization of the trapped wavepacket in
a way typical of the multiphoton regime: fragments of the
wavepacket are torn off and sent in all directions. It is how the
first, main family of the rings in Figs \ref{fig3} is formed. The
torn-off pieces, being of a considerable size, are scattered on
the other minimum of the KH well, which results in a formation of
the second family of rings, centered about
$2\varepsilon_0/\omega^2$. Both families of the rings can
interfere. In particular this is visible for the parts moving
along the \emph{x}-axis: a pattern structure is built after a few
cycles between the two minima of the KH well.

In the case of a sine pulse with a trapezoid envelope the
mechanism of tearing is similar as described above. An important
difference is that it takes a few cycles for the packet to become
broad enough so that it reaches the minima of the KH well and can
be ionized by higher terms of the Fourier expansion. After that
period the packet is rescattered on both minima of the KH well.
Both families of rings appear simultaneously.

Our observations in two dimension allowed us to complete the
interpretation of one dimensional results: a very similar
interference structure in the form of numerous subpeaks between
the classical turning points can also be seen in the results of
one-dimensional computations (see Fig. \ref{fig7}). The effect is
significantly reduced if the well is made more shallow or if it is
replaced by a soft-core potential.

\section{Conclusions}
We have presented a detailed dynamical picture of two-dimensional
photodetachment in very strong fields in the parameters range not
explored before. Due to the rapidly varying binding potential, for
some model pulses one can observe tearing smaller pieces off the
main packet. The latter pieces can themselves be torn into
fragments, giving rise to new spatial structures of the
wavepacket, possibly interfering with the old ones. We have
demonstrated those elements of the dynamics that are due to the
binding potential being short-range. The behavior of the packet
described above is to be contrasted with the packet's dynamics in
the case of a smooth binding potential, in which a connected
structure is trapped in KH well and possibly push out by the
magnetic force.

\newpage
\begin{figure}
\caption{The KH averaged potential for a quantum well potential
($a=1$ a.u., $V_0=2$ a.u.) and a cosine pulse ($\varepsilon_0=15$
a.u., $\omega=1$ a.u.). On the left – the KH potential and the
ground state wavefunction with energy $E_1=-0.0192$ a.u. On the
right – the wavefunction of the only excited state ($E_2=-0.0158$
a.u.).}
\label{fig1}
\end{figure}

\begin{figure}
\caption{The population of the initial state. The pulse parameters
are $\varepsilon_0=15$ a.u., $\omega =1$ a.u. and the well
potential $a=1$ a.u., $V_0=2$ a.u.} \label{fig2}
\end{figure}

\begin{figure}
\caption{The modulus square of the wavefunction after $1T$, $2T$,
$3T$, $4T$, $4.8T$, $5T$, $10T$ without the dipole approximation
and after $10T$ in the dipole approximation. The distance between
the centers in the $x$ direction (horizontal) of the families of
the rings is 30 a.u. The atomic and field parameters are the same
as in Fig. \ref{fig2}.} \label{fig3}
\end{figure}

\begin{figure}
\caption{The solution of the Newton equations for an unbound
electron without the dipole approximation ($x(t)$ on the upper
plot and $y(t)$ on the lower one). The trajectory in the $x$
direction in the dipole approximation coincides with that without
this approximation in the scale of the picture. A cosine pulse is
assumed, with the intensity $\varepsilon_0=15$ a.u. and frequency
$\omega =1$ a.u. One can see fast oscillations in the $x$
direction and the magnetic drift in the \emph{y} direction.}
\label{fig4}
\end{figure}

\begin{figure}
\caption{The mean values of $x$ and $y$ in the quantum simulation
for the pulse parameters as in Fig. \ref{fig4}. One can see the
slow drift in the \emph{x} direction and the magnetic drift in the
\emph{y} direction.}
\label{fig5}
\end{figure}

\begin{figure}
\caption{Modulus square of the wavefunction after $6T$. One can
distinguish fragments thrown away after full cycles from those
thrown away in the middle of each cycle (see inset). The
parameters of a rectangular cosinusoidal pulse are
$\varepsilon_0=20$ a.u., $\omega =1$ a.u. The shift of the
probability density towards the region of $y>0$ is well visible.}
\label{fig6a}
\end{figure}

\begin{figure}
\caption{As in Fig. \ref{fig6a} but in the dipole approximation.}
\label{fig6b}
\end{figure}

\begin{figure}
\caption{The main picture: a one-dimensional wavepacket for a
rectangular pulse with the parameters as in Fig. \ref{fig3} and
the well potential chosen to match the energy of only state
$E_1=-0.832$ a.u. The small pictures: upper wavepacket, the same
field but for a soft-core atom \cite{4}; lower wavepacket, the
well potential four times less deep than in the main picture (the
maxima on the small pictures have the height about 0.0004 a.u.).}
\label{fig7}
\end{figure}

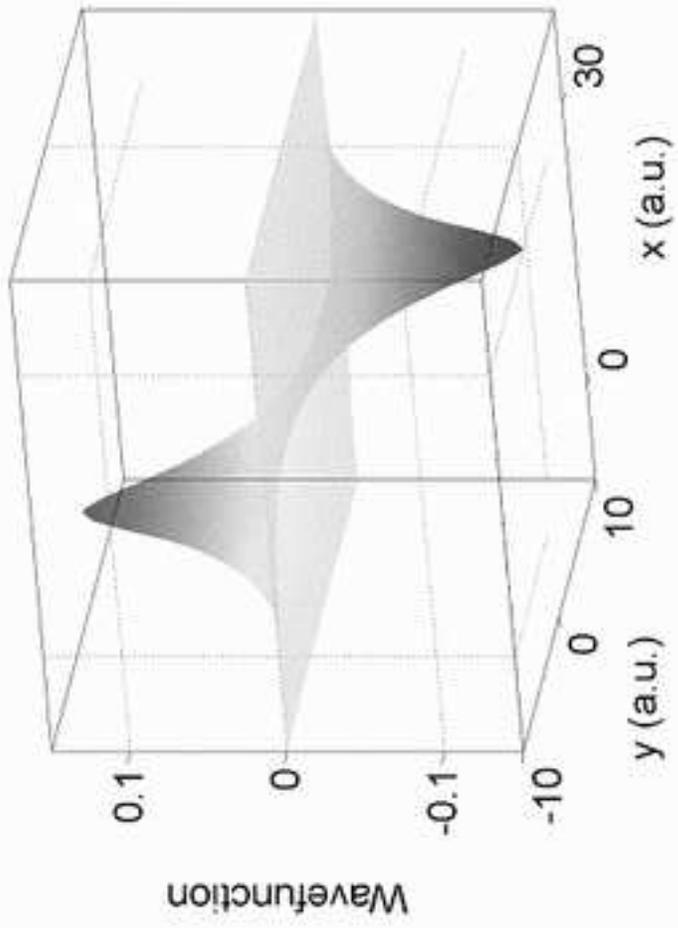
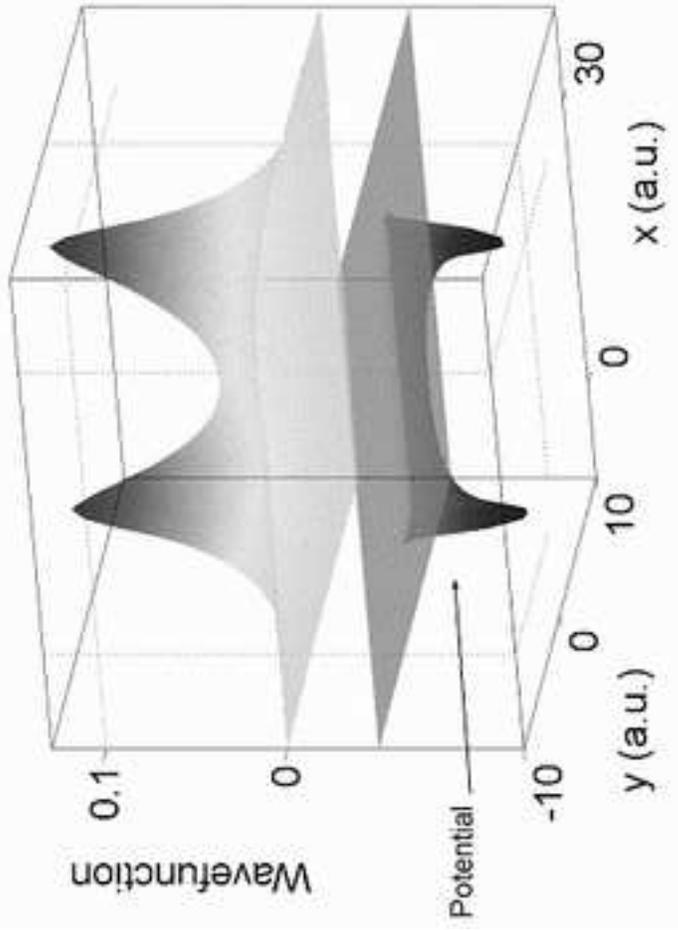
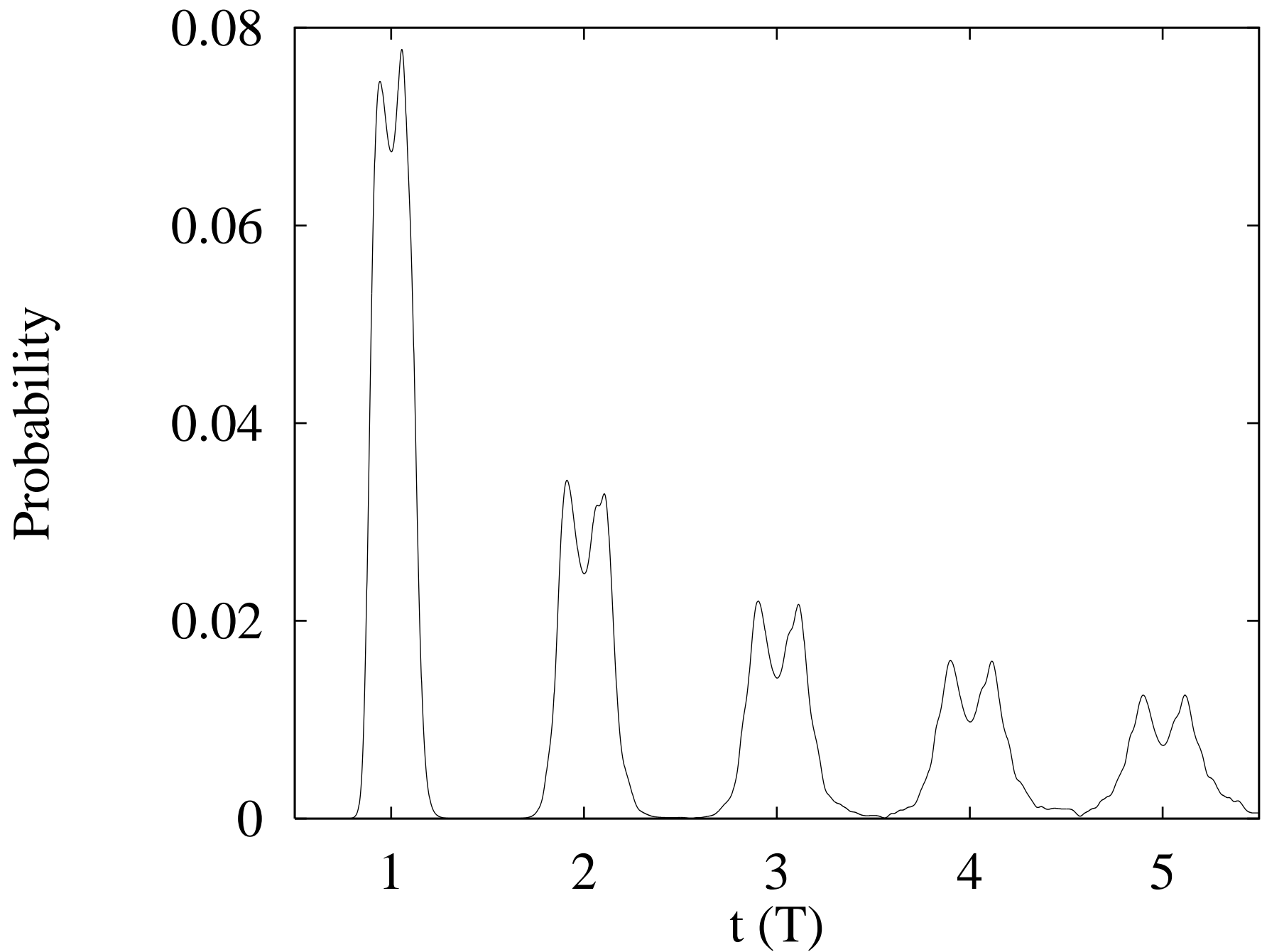
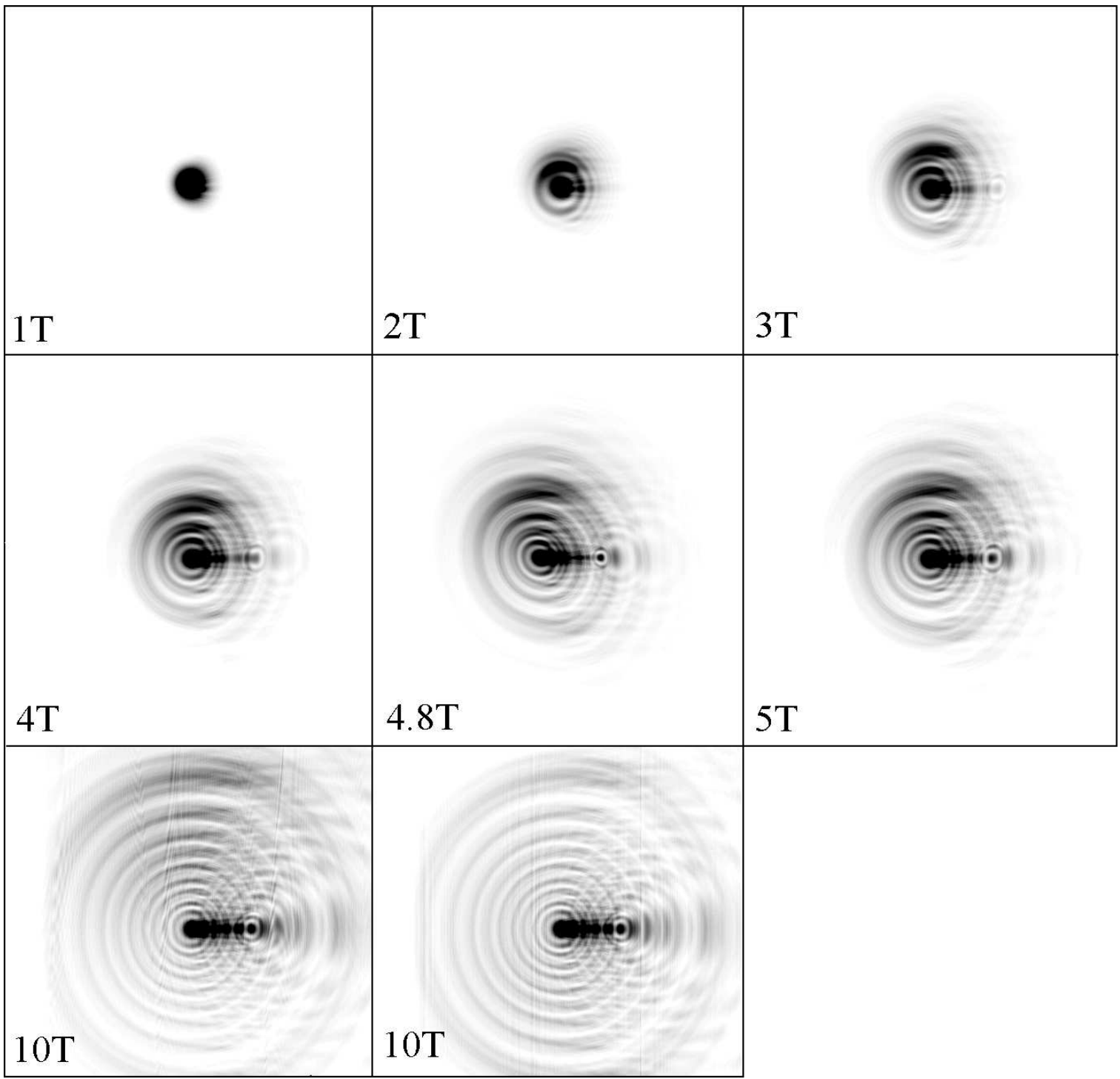
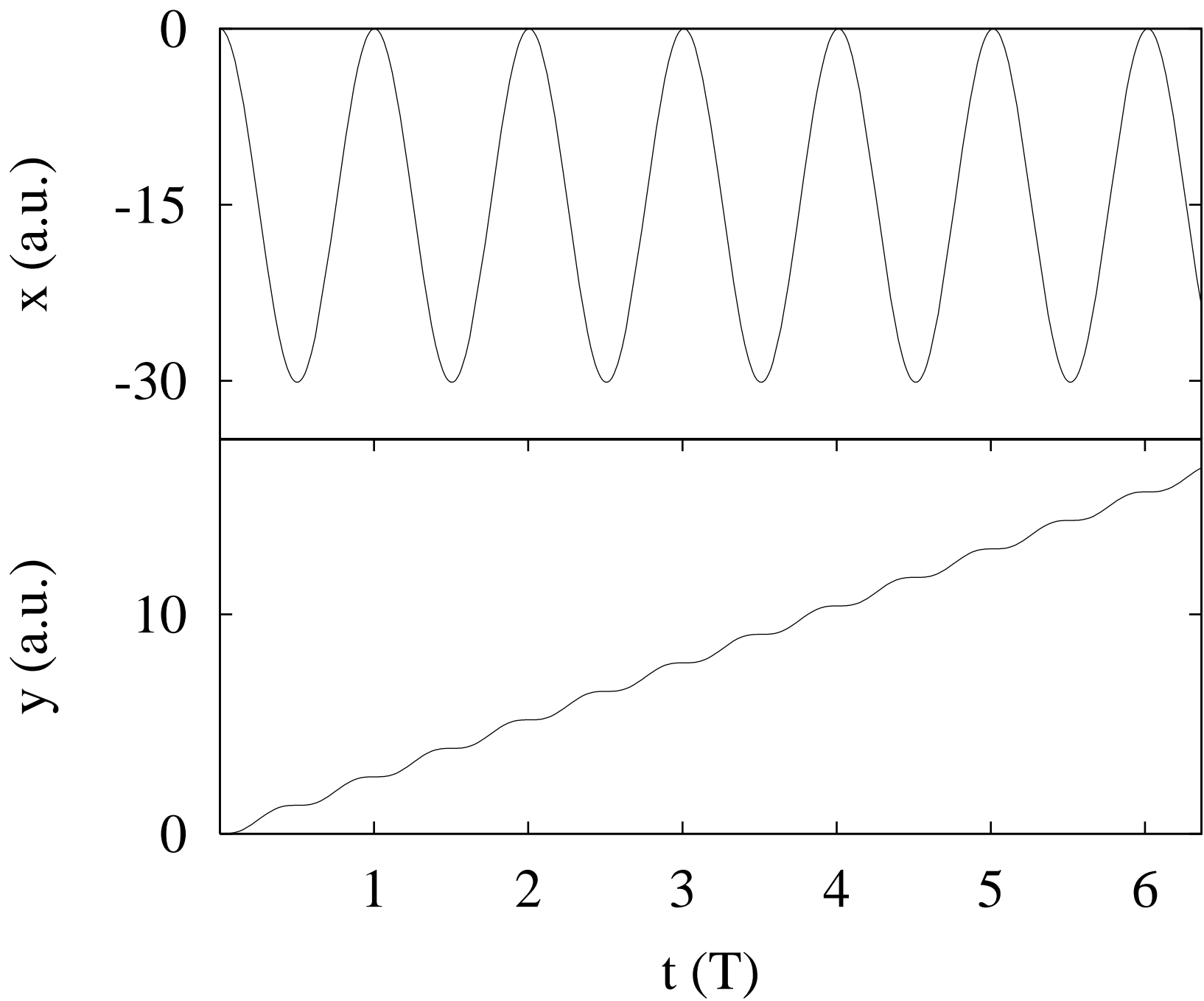
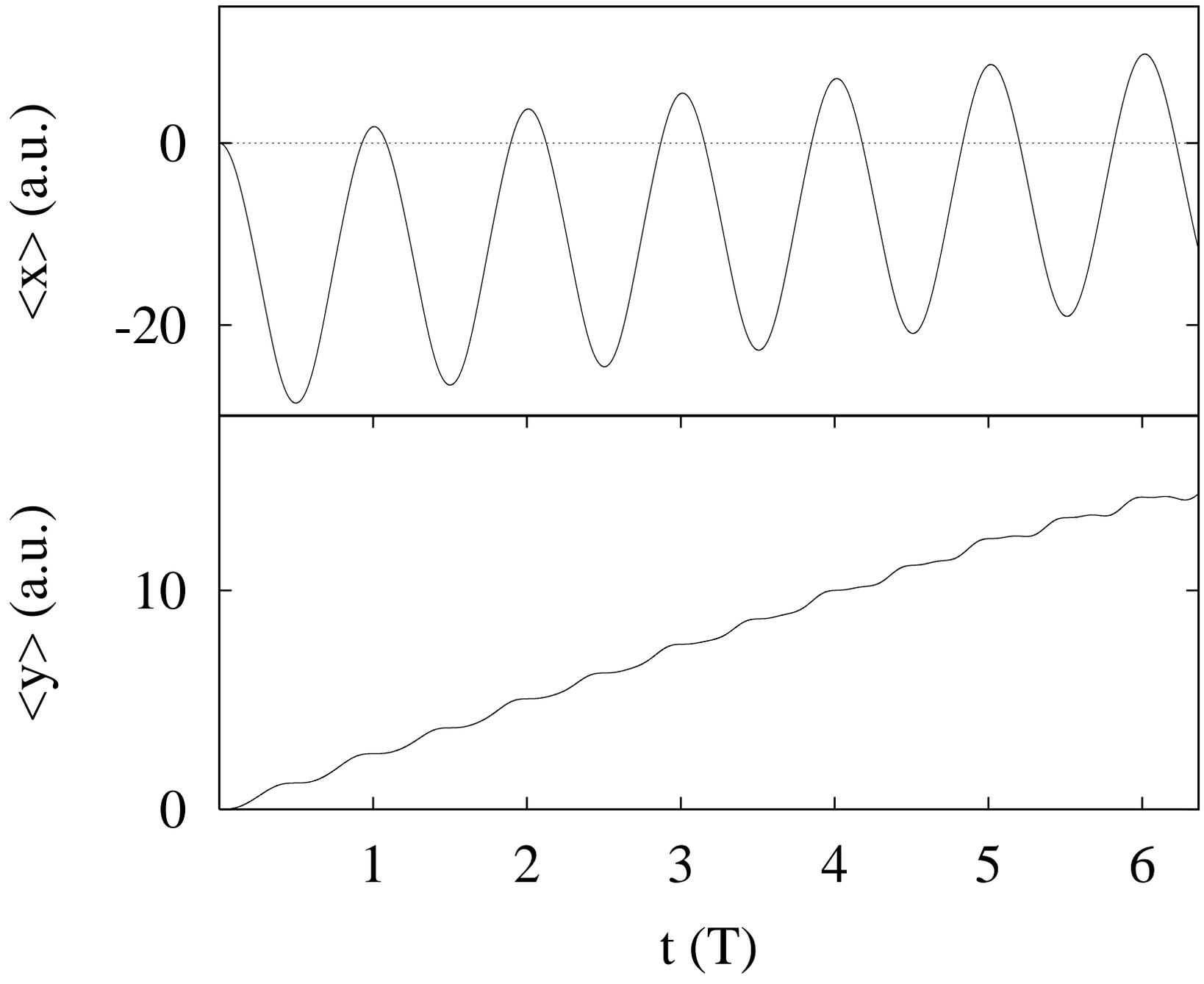
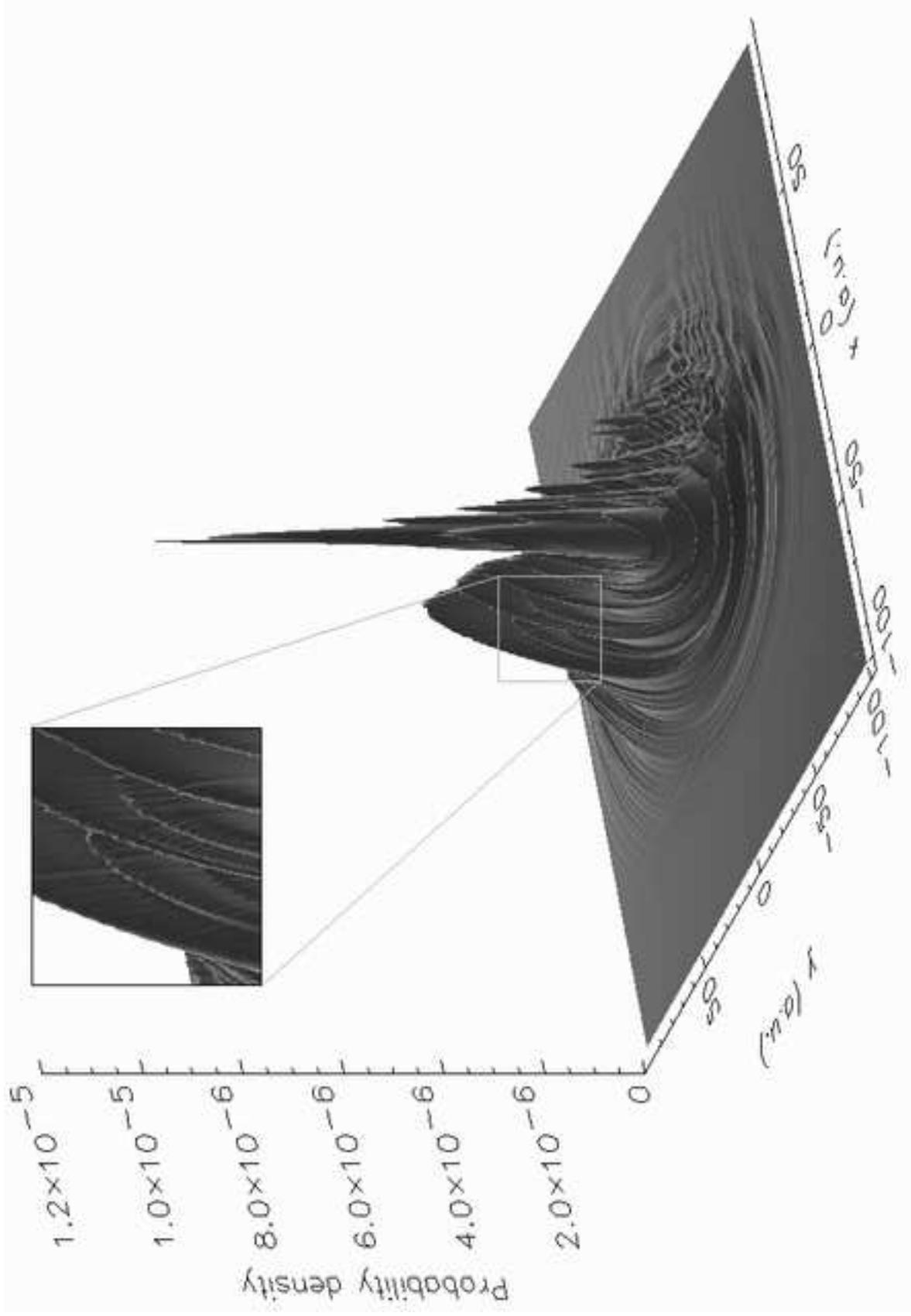
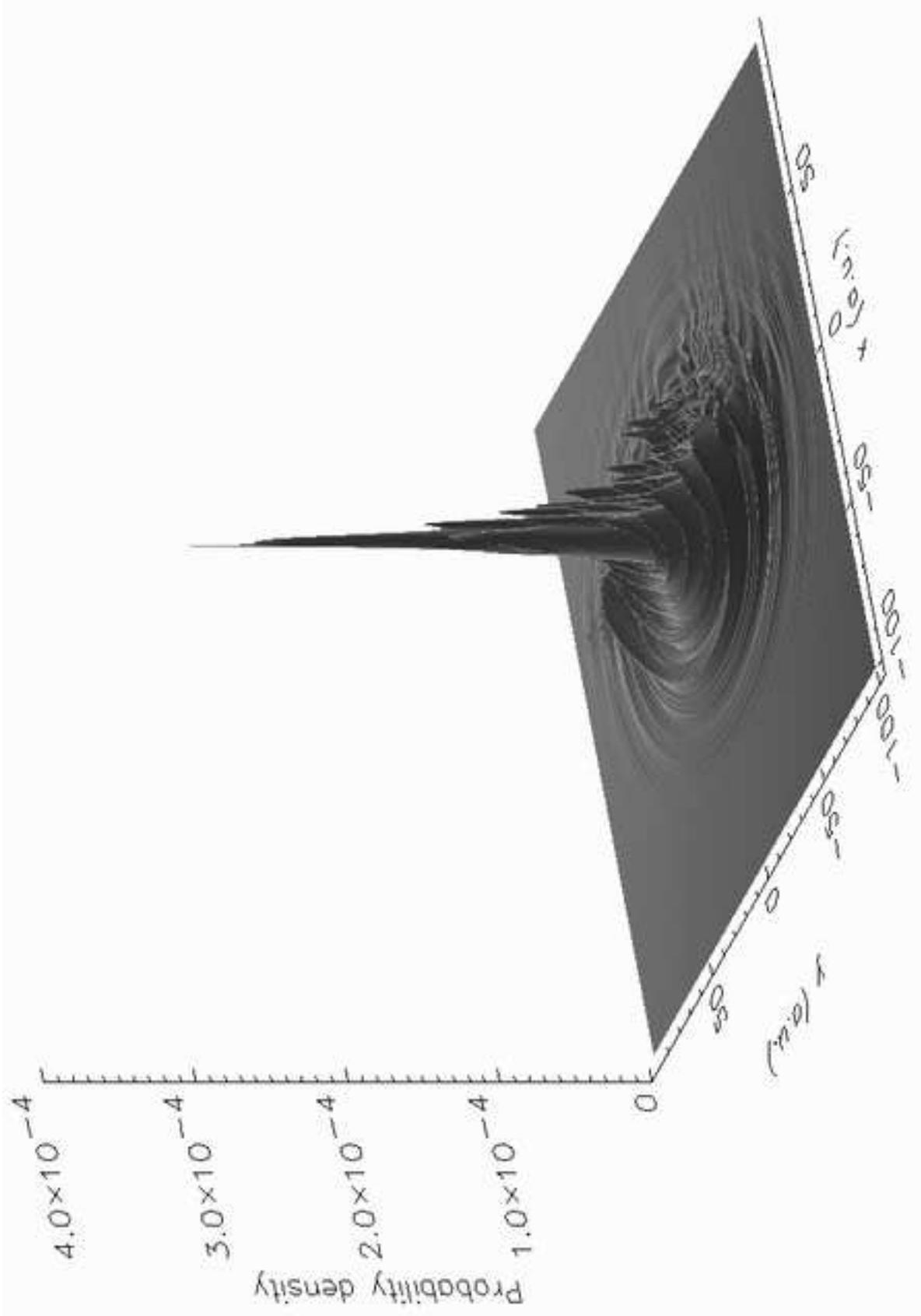
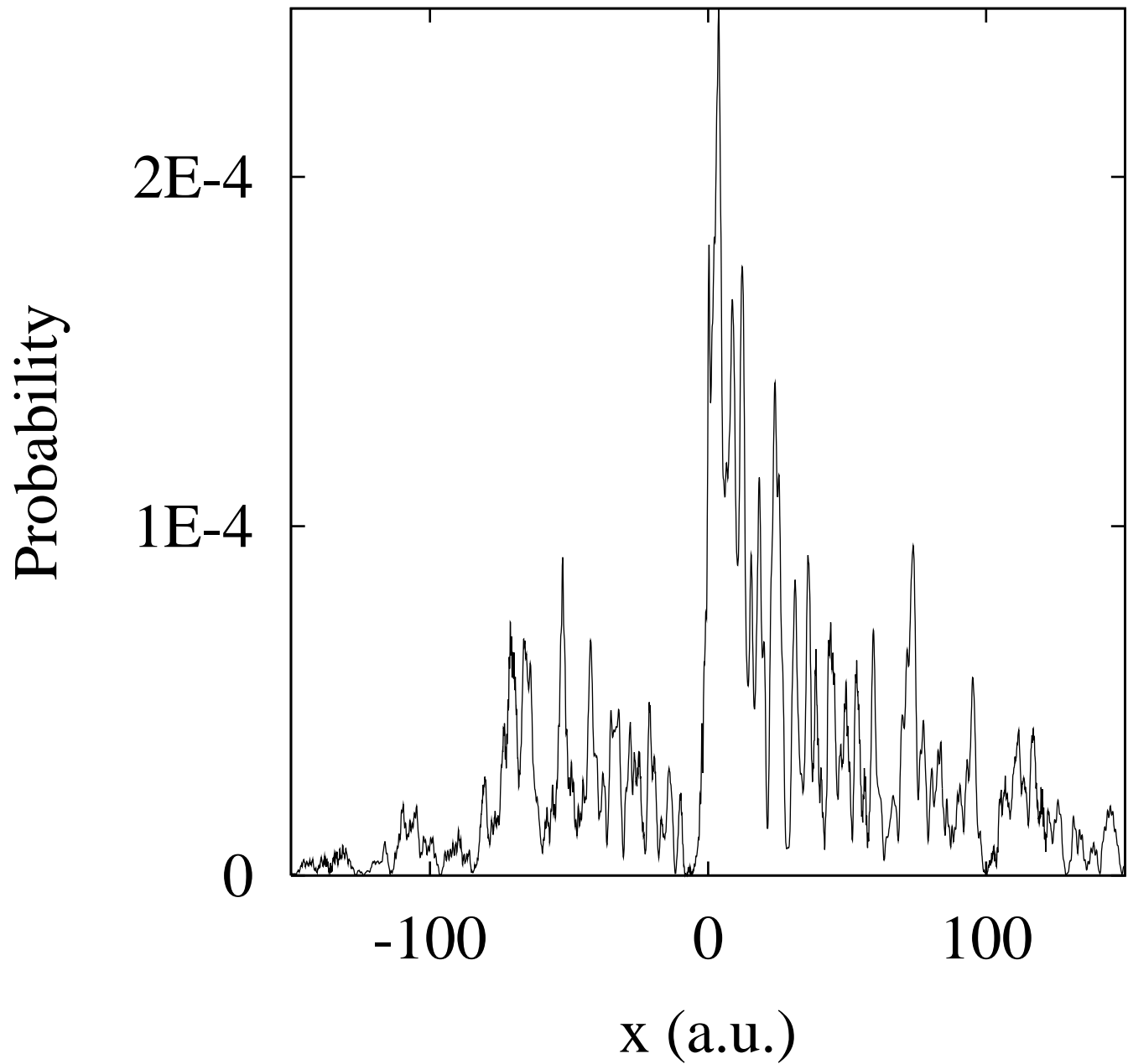
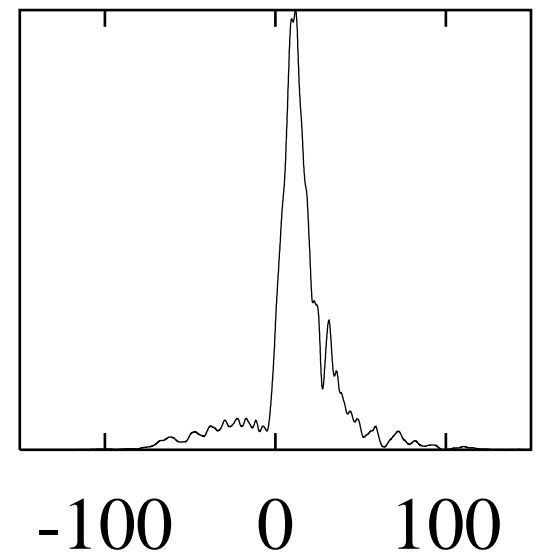
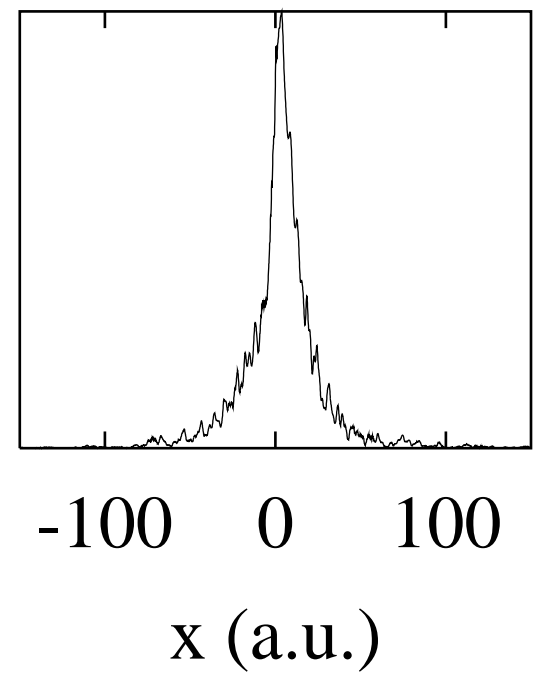
\end{document}